\documentclass[a4paper,11pt]{article}
\usepackage{pos}

\usepackage{natbib}
\usepackage{cleveref}
\usepackage{graphicx}
\usepackage{floatrow}

\usepackage{multirow}

\newcommand{\af}[1]{\textcolor{blue}{#1}}

\title{Deflated Multigrid Multilevel Monte Carlo}
\ShortTitle{Deflated Multigrid Multilevel Monte Carlo}

\author[a]{Andreas Frommer}
\author*[a]{Gustavo Ramirez-Hidalgo}

\affiliation[a]{Department of Mathematics, Bergische Universit{\"a}t Wuppertal, 42097 Wuppertal, Germany}

\emailAdd{g.ramirez@math.uni-wuppertal.de}

\abstract{

In lattice QCD, the trace of the inverse of the discretized Dirac operator appears in the disconnected fermion loop contribution to an observable. As simulation methods get more and more precise, these contributions become increasingly important. Hence, we consider here the problem of computing the trace $\mathrm{tr}(D^{-1})$, with $D$ the Dirac operator. The Hutchinson method, which is very frequently used to stochastically estimate the trace of a function of a matrix, approximates the trace as the average over estimates of the form $x^{H} D^{-1} x$, with the entries of the vector $x$ following a certain probability distribution. For $N$ samples, the accuracy is $\mathcal{O}(1/\sqrt{N})$. In recent work, we have introduced multigrid multilevel Monte Carlo: having a multigrid hierarchy with operators $D_{\ell}$, $P_{\ell}$ and $R_{\ell}$, for level $\ell$, we can rewrite the trace $\mathrm{tr}(D^{-1})$ via a telescopic sum with difference-levels, written in terms of the aforementioned operators and with a reduced variance. We have seen significant reductions in the variance and the total work with respect to exactly deflated Hutchinson. In this work, we explore the use of exact deflation in combination with the multigrid multilevel Monte Carlo method, and demonstrate how this leads to both algorithmic and computational gains.

}

\FullConference{%
The 39th International Symposium on Lattice Field Theory,\\
8th-13th August, 2022,\\
Rheinische Friedrich-Wilhelms-Universität Bonn, Bonn, Germany
}

\begin{document}
\maketitle

\section{Introduction}

In lattice quantum chromodynamics (LQCD), the extraction of some observables requires the computation of matrix traces \citep{gambhir2017disconnected} of the form $\mathrm{tr}(f(D))$ with $D$ the Dirac operator on the lattice, and $f(D) = (\Gamma D)^{-1}$ with $\Gamma$ some Dirac structure e.g.\ $\Gamma = \Gamma_{5}$. This appears, for example, in the calculation of disconnected diagrams \citep{gambhir2017disconnected}.

There are deterministic algorithms for the approximate computation of such traces, among which falls hierarchical probing \cite{Stathopoulos2013,LaeuchliStathopoulos2020}. From the stochastic side, on which we focus in this work, a well known method is the use of the Hutchinson estimator \cite{Hutchinson90}
\begin{equation}
\mathrm{tr}(f(A)) \approx \frac{1}{N} \sum_{i=1}^{N} (x^{(i)})^{H} f(D) x^{(i)}.
\label{eq:Hutchinson_estimator}
\end{equation}

The entries $x_j$ in the vectors $x^{(i)}$ in \cref{eq:Hutchinson_estimator} are random identically and independently distributed (i.i.d.) with expected values $\mathbb{E}[x_i] = 0$ and $\mathbb{E}[x_ix_j] = \delta_{ij}$. The variance $\mathbb{V}[x^{H} f(D) x ]$, in terms of the entries of the matrix $f(D)$, is determined by the probability distribution (p.d.f.) used for drawing the components of the vector $x$.

With $\mathbb{V}(x^{H} f(D) x)$ having a theoretical value that depends on $f(D)$ and the p.d.f.\ used in drawing $x$, the variance of the Hutchinson estimator in \cref{eq:Hutchinson_estimator} decreases as $\frac{1}{\sqrt{N}}$. When the accuracy required in the computation of $\mathrm{tr}(f(D))$ is quite high, the Hutchinson method becomes too costly. There are multiple variance reduction techniques that can be applied to the Hutchinson estimator, in particular deflation \cite{Gambhir_2017}, which we discuss in \cref{sect:deflated_hutchinson}.

A new variance reduction technique recently developed in \cite{frommer2022multilevel}, which we brief\/ly describe in \cref{sect:mgmlmc}, uses a multilevel approach via a multigrid hierarchy in order to ``distribute" the variance over different multigrid levels, and with this offloading some (and in some cases most) of the computational work to coarser levels.

In this work, we discuss the use use of exact deflation on the multigrid multilevel Monte Carlo method from \cite{frommer2022multilevel}. We focus on $f(D) = D^{-1}$, with $D$ the Wilson-Schwinger operator coming from a discretization of the (1+1)-dimensional Schwinger model on the lattice.

\section{Deflated Hutchinson}
\label{sect:deflated_hutchinson}

If $\mathbb{Z}_{2}$ noise is used for drawing the random vectors used in the Hutchinson estimator, then \cite{DongLiu1993,wilcox2000noise}
\begin{equation}
\mathbb{V}(x^{H} D^{-1} x) = \frac{1}{2} \| \mathrm{offdiag}(D^{-1} + D^{-T}) \|^{2}_{F}.
\label{eq:variance_in_terms_of_frob_norm}
\end{equation}

Now, for a given matrix $A \in \mathbb{C}^{n \times n}$, there is a connection between the Frobenius norm of $A$ and its singular values
\begin{equation}
\| A \|^{2}_{F} = \sum_{i=1}^{n} \sigma_{i}^{2} \, \Rightarrow \, \| \mathrm{offdiag}(A) \|^{2}_{F}= \sum_{i=1}^{n} \sigma_{i}^{2} - \sum_{i=1}^{n} | A_{ii} |^{2}.
\label{eq:relation_frob_norm_and_singvalues}
\end{equation}

In LQCD simulations, where the smallest singular values of $D$ are typically very small---and those of $D^{-1}$ are thus large---, the last term in \cref{eq:relation_frob_norm_and_singvalues} can be discarded, and we see that the smallest singular modes of $D$ are the ones dominating the variance. Deflating those modes can then lead to a significant reduction in the variance of the Hutchinson estimator.

Exactly deflated Hutchinson \cite{Gambhir_2017} has been extensively used in LQCD, e.g.\ \cite{alexandrou2019proton,alexandrou2020complete,davoudi2021nuclear}. With the singular value decomposition (SVD) \cite{trefethen1997numerical} of the Dirac operator, $D = U \Sigma V^{H}$ and its inverse $D^{-1} = V \Sigma^{-1} U^{H}$, exact deflation is done via the orthogonal projector
\begin{equation}
\Pi = U_{k} U_{k}^{H},
\label{eq:exact_deflation_projector}
\end{equation}
where the columns of $U_{k}$ contain the $k$ largest right singular vectors of $D^{-1}$, i.e., the $k$ smallest left singular vectors of $D$. The trace $\mathrm{tr}(D^{-1})$ can then be split as
\begin{equation}
\mathrm{tr}(D^{-1}) = \mathrm{tr}( D^{-1}(I-\Pi) ) + \mathrm{tr}( D^{-1}\Pi ).
\label{eq:splitting_defl_trace}
\end{equation}

From the discussion following \cref{eq:variance_in_terms_of_frob_norm}, the first term in \cref{eq:splitting_defl_trace} will have a reduced variance, so we can compute that term via the Hutchinson estimator investing less estimates. For the second term, we can write
\begin{equation}
\mathrm{tr}( D^{-1}\Pi ) = \mathrm{tr}( D^{-1} U_{k} U_{k}^{H} ) = \mathrm{tr}( U_{k}^{H} D^{-1} U_{k} ) = \mathrm{tr}( U_{k}^{H} V_{k} \Sigma_{k}^{-1} ),
\label{eq:small_trace_term_defl_Hutchinson}
\end{equation}
where the last step requires a pre-computation of the smallest $k$ left singular vectors of $D$ to high accuracy.

Better algorithms are typically known for eigenvalue problems, compared to singular value problems. The Dirac operator being $\Gamma_{5}$-Hermitian \cite{Gattringer_and_Lang_book}, i.e.\ $( \Gamma_{5} D )^{H} = \Gamma_{5} D$, we can extract the singular vectors of $D$ from the eigenvectors of $Q = \Gamma_{5} D$
\begin{equation}
Q = X \Lambda X^{H} \Rightarrow D = (\Gamma_{5} X \mathrm{sign}(\Lambda)) \mathrm{abs}(\Lambda) X^{H},
\end{equation}
with $X$ the eigenvectors of $Q$, $\Lambda$ the diagonal matrix containing the eigenvalues of $Q$, $U = \Gamma_{5} X \mathrm{sign}(\Lambda)$ the left singular vectors of $D$ and $V = X$ its right singular vectors. The matrix $\mathrm{abs}(\Lambda)$, which corresponds to the singular values of $D$, is simply the element-wise absolute value of $\Lambda$, and $\mathrm{sign}(\Lambda)$ is such that $\Lambda = \mathrm{sign}(\Lambda) \mathrm{abs}(\Lambda)$.

\section{Multigrid Multilevel Monte Carlo}
\label{sect:mgmlmc}

The variance reduction technique introduced in \cite{frommer2022multilevel} makes use of a multigrid hierarchy. We explain now what a multigrid hierarchy is, its relation to inexact deflation, and its use in multigrid multilevel Monte Carlo.

\subsection{Multigrid}

Already with the basic Hutchinson estimator, with or without deflation, LQCD simulations need to solve linear systems of equations of the form $Dx = b$. The Dirac operator $D$ is typically ill-conditioned, with a spectrum that makes it hard for traditional methods such as Krylov-based iterative solvers to find the solution $x$  with tolerable effort. In current LQCD simulations, multigrid solvers \cite{ruge1987algebraic,briggs2000multigrid} are the state of the art when dealing with $Dx=b$ \cite{osbornpos10,frommer2014adaptive,brower1987multigrid,
Frommer:2013fsa}.

In a two-level multigrid solver, a \textit{smoother} is applied at the original grid (i.e.\ the finest level), followed by a \textit{coarse-grid correction}. The smoother, which typically consists of a few iterations of a method such as Gauss-Seidel, Jacobi or GMRES, removes high-frequency components of the error, and the coarse-grid correction serves as a complement to the smoother, in charge of dealing with those modes of the error that have not been dealt with by the smoother. In a more than two-level multigrid method, the coarse grid correction is obtained by applying the two-level method recursively. The operator at level $\ell$ of the resulting multigrid hierarchy is labeled as $D_{\ell}$. There is an operator that allows the transfer of data from level $\ell$ to $\ell+1$, known as the restriction operator $R_{\ell}$; moving data in the opposite direction is done via the interpolation operator $P_{\ell}$.

Mathematically, one iteration of two-level multigrid can be described by the sequence
\begin{equation}
\begin{split}
r & \leftarrow b - D_{1} x \\
x & \leftarrow x + S^{(\nu_{1})}_{1} r \ \ \ \ \ (\text{pre-smoothing, $\nu_1$ iterations}) \\
r & \leftarrow b - D_{1} x \\
x & \leftarrow x + P_{1} D_{2}^{-1} R_{1} r \ \ \ \ \ (\text{coarse-grid correction}) \\
r & \leftarrow b - D_{1} x \\
x & \leftarrow x + S^{(\nu_{2})}_{1} r  \ \ \ \ \ (\text{post-smoothing, $\nu_2$ iterations})
\end{split}
\label{eq:two_level_mg_method}
\end{equation}
where $r$ is the \textit{residual}, and the operator $S^{(\nu)}_{\ell}$ is the smoother at level $\ell$, with $\nu$ the number of iterations of the smoother. There are different ways of constructing the coarse-grid operator $D_{2}$, e.g.\ the Petrov-Galerkin approach $D_{2} = R_{1}D_{1} P_{1}$.

If the matrix $D_{2}$ is still too large, the application $D_{2}^{-1}(R_{1}r)$ in \cref{eq:two_level_mg_method} can be further computed via another two-level method. This can be put up in a recursive manner, leading to a multilevel multigrid solver.

The interpolator $P_{\ell}$ can be built in a geometric or algebraic way. In the former, $P_{\ell}$ is based on the geometry of the lattice and information of an underlying infinite-dimensional operator prior to discretization . In the latter, $P_{\ell}$ is rather constructed via the information contained in the matrix $D_{\ell}$. Due to the random nature of the gauge links in LQCD, algebraic multigrid is necessary to have an effective linear solver. Furthermore, the aggregation-based construction of $P_{\ell}$ from $D_{\ell}$ in lattice QCD relies on a concept known as \textit{local coherence} \cite{luscher2007local}, which states that many low modes of $D_{\ell}$ can be approximately obtained from just a few low modes of the same operator, by looking at the local behaviour of those few modes. The construction of the multigrid hierarchy that we follow here is the one presented in \cite{Rottmann:2016gmm}.

\subsection{Inexact Deflation and Multigrid Multilevel Monte Carlo}

The orthogonal projector in \cref{eq:exact_deflation_projector} is built using exact singular vectors $u_i$. If for computational efficiency these are only computed approximately, the computation of $\mathrm{tr}(D^{-1}\Pi) = \mathrm{tr}(D^{-1}UU^*) = \mathrm{tr}(U^*D^{-1}U)$ requires $k$ extra inversions in \cref{eq:splitting_defl_trace}. A cheaper alternative is to use inexact deflation as in \cite{romero2020multigrid}. For this, the projector
\begin{equation}\label{eq:projector_inex_defl}
    \Pi = U_{k} ( V_{k}^{H} D U_{k} )^{-1} V_{k}^{H} D
\end{equation}
is used. This \textit{oblique}  projector splits the trace, using $D^{-1} = (I-\Pi)D^{-1} + \Pi D^{-1}$, as
\begin{equation}
\mathrm{tr}(D^{-1}) = \mathrm{tr}( D^{-1} - U_{k} ( V_{k}^{H} D U_{k} )^{-1} V_{k}^{H} ) + \mathrm{tr}( ( V_{k}^{H} D U_{k} )^{-1} V_{k}^{H} U_{k} )
\label{eq:projector_inexact_defl}
\end{equation}
The second term in \cref{eq:projector_inexact_defl}  now requires only $k$ matrix-vector multiplications with $D$ acting on the $k$ deflation vectors $u_i$ and the inversion of the small $k \times k$ matrix $V_{k}^{H} D U_{k}$, but no solves with the large matrix $D$.  The first term in \cref{eq:projector_inexact_defl} is expected to have a reduced variance, similar to the case of the orthogonal projector \eqref{eq:exact_deflation_projector} in exact deflation.

In the multigrid context we know from local coherence \cite{luscher2007local}, that the range of the projection $P_{1}$ is composed of many approximate low modes of $D_{1}$. This fact is used in \cite{romero2020multigrid} to construct inexact deflation with the projector $\Pi_1 = P_1(P_1^HDP_1)^{-1}P_1^H D = P_1D_2^{-1}P_1^H D$.

Based on the general multilevel Monte Carlo approach \cite{Giles2015},
the multigrid multilevel Monte Carlo method proposed in \cite{frommer2022multilevel} applies this construction recursively to obtain the splitting (note that $P_\ell^HP_\ell = I$)
\begin{equation}
\mathrm{tr}(D^{-1}) = \sum_{\ell=1}^{L-1} \mathrm{tr}\left(D_\ell^{-1}- {P}_{\ell} D_{\ell+1}^{-1} {P}^{H}_{\ell}\right) + \mathrm{tr}(D_L^{-1}),
\label{eq:mgmlmc}
\end{equation}
where $\ell$ runs over the different levels in the multigrid hierarchy, with $\ell=L$ being the coarsest level, and $D_1 = D$. The goal in \cref{eq:mgmlmc} is to have a sequence of difference-level operators $M_{\ell} := D_\ell^{-1}- {P}_{\ell} D_{\ell+1}^{-1} {P}^{H}_{\ell}$ with reduced variance when computing their traces, and a last term $\mathrm{tr}(D_{L}^{-1})$ with large variance but cheap to compute.

\begin{figure}[h]
\caption{Cost versus mass parameter $m$ in the computation of the trace $\mathrm{tr}(D^{-1})$, with $\texttt{eps}=10^{-3}$, for a (1+1)-dimensional Wilson-Schwinger operator $D$ and a lattice of size $128^{2}$. A four-level multigrid hierarchy is employed. Figure taken from \cite{frommer2022multilevel}.}
\centering
\includegraphics[width=.55\textwidth]{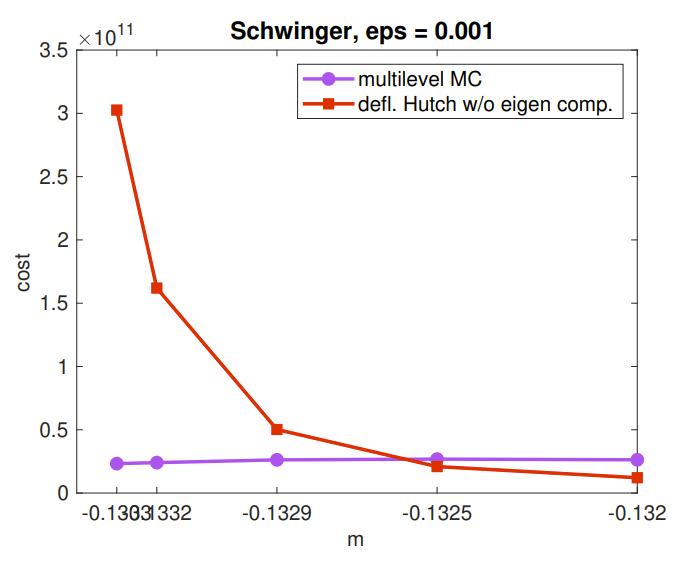}
\label{fig:results_mlmlc_paper}
\end{figure}

The expression in \cref{eq:mgmlmc} has been used in \cite{frommer2022multilevel} to compute $\mathrm{tr}(D^{-1})$ in the case of the (1+1)-dimensional Schwinger model, in particular. Results for a $128^2$ lattice are displayed here in \cref{fig:results_mlmlc_paper}, where $m$ is the mass parameter of the Dirac operator, \texttt{cost} is given in FLOPS, and \texttt{eps} is the relative tolerance used in the stopping condition.
The multigrid multilevel Monte Carlo method presented in \cite{frommer2022multilevel} displays insensitivity to conditioning, in the example illustrated here in \cref{fig:results_mlmlc_paper}, and outperforms a highly tuned exactly deflated Hutchinson by a factor of around 12 for the most ill-conditioned case (i.e.\ for the smallest value of $m$). Note that we did not include the work for computing the eigenvectors used in exactly deflated Hutchinson in fig.~\ref{fig:results_mlmlc_paper}.

\section{Deflated Multigrid Multilevel Monte Carlo}

In \cref{fig:results_mlmlc_paper}, deflated Hutchinson outperforms multigrid multilevel Monte Carlo for $m=-0.132$ and $m=-0.1325$. We can revert that via two improvements:

\begin{itemize}
    \item skipping a level
    \item applying exact deflation on every difference level in multigrid multilevel Monte Carlo
\end{itemize}

\subsection{Skipping a level}

From the numerical experiments performed in \cite{frommer2022multilevel} for the Schwinger model, the number of nonzero elements in $D_{1}$ is approximately the same as in $D_{2}$. This renders the second level almost as expensive as the first one, which we take into account in this section.

Assuming a four-level multigrid hierarchy, and skipping the second level in the trace decomposition in \cref{eq:mgmlmc}, we can then write
\begin{equation}
\mathrm{tr}(D^{-1}) = \mathrm{tr}(D_1^{-1} - P_{1} P_{2} A_{3}^{-1} P^H_{2} P^H_{1} ) +\mathrm{tr}( A_{3}^{-1} -  P_{3} A_{4}^{-1} P^H_{3}  ) + \mathrm{tr}( A_{4}^{-1} )
\label{eq:skip_level}
\end{equation}

The first term in \cref{eq:skip_level} might see an increased variance with respect to the first or second terms in the summation in \cref{eq:mgmlmc}, but we might see a gain in total cost due to avoiding inversions with $D_{2}$ in the multilevel trace expansion.

\subsection{Exact deflation on difference levels}
\label{sect:exact_defl_diff_levels}

Following the discussion in \cref{sect:deflated_hutchinson}, we can apply a similar exact deflation  approach on each difference level in multigrid multilevel Monte Carlo. For this, we compute the largest singular vectors for the difference-level operator
\begin{equation}
    M_{\ell} := D_\ell^{-1}- {P}_{\ell} D_{\ell+1}^{-1} {P}^{H}_{\ell}.
\end{equation}

We can reduce the computation of singular triplets to eigenpairs in a way to what we outlined in \cref{sect:deflated_hutchinson}. We can construct Hermitian difference-level operators using the relations
\begin{equation}
    \Gamma_{5}^{\ell} P_{\ell} = P_{\ell} \Gamma_{5}^{\ell+1}, \ \ P^{H}_{\ell} \Gamma_{5}^{\ell} = \Gamma_{5}^{\ell+1} P^{H}_{\ell},\ \ (\Gamma_5^\ell)^H = \Gamma_5^\ell, \ \ (\Gamma_5^\ell)^H\Gamma_5^\ell =  I,
\end{equation}
which come from the ``spin-preserving'' algebraic multigrid construction discussed in \af{\cite{frommer2014adaptive}} and
which imply
\[
Q_\ell := \Gamma_5^\ell D_\ell = D_\ell^H \Gamma_5^\ell = Q_\ell^H.
\]
With these relations at hand, we obtain a Hermitian operator which we can use for the indirect extraction of the singular vectors of $M_{\ell}$:
\begin{equation}
    J_{\ell} = M_{\ell} \Gamma_{5}^{\ell} =  D^{-1}_{\ell} \Gamma_{5}^{\ell} - P_{\ell}D^{-1}_{\ell+1}P^{H}_{\ell} \Gamma_{5}^{\ell} = Q^{-1}_{\ell} -  P_{\ell} D^{-1}_{\ell+1} \Gamma_{5}^{\ell+1} P^{H}_{\ell} = Q^{-1}_{\ell} - P_{\ell} Q^{-1}_{\ell+1} P^{H}_{\ell}.
\end{equation}
Then, in the same way as in \cref{sect:deflated_hutchinson}, we extract singular vectors of $M_{\ell}$ via eigenvectors of $J_{\ell}$

\begin{equation} \label{eq:hermitian_diff_lev_op}
    J_{\ell} = X \Lambda X^{H} \Rightarrow M_{\ell} = X \Lambda X^{H} \Gamma^{\ell}_{5} \Rightarrow U = X \mathrm{sign}(\Lambda), S = \mathrm{abs}(\Lambda), V = \Gamma_{5} X
\end{equation}

\subsection{Results}

We have implemented the computation of $\mathrm{tr}(D^{-1})$ via deflated multigrid multilevel Monte Carlo in Python\footnote{The code can be found \href{https://github.com/Gustavroot/DeflatedMLMC_Schwinger}{here}.}. We have performed a numerical test with the exact same Schwinger matrix used in \cite{frommer2022multilevel}. We take $m_{0} = -0.1320$ here, and we run on a single core of a node with 44 cores Intel(R) Xeon(R) CPU E5-2699 v4 @ 2.20GHz. We seek the trace with a relative tolerance of $10^{-4}$, with a relative residual norm of $10^{-12}$ for the linear solves on each sample. When using exact deflation, the relative tolerance in the eigensolver is $10^{-9}$, and $10^{-1}$ for inexact deflation. When calling the eigensolver on $J_{\ell}$, see \cref{eq:hermitian_diff_lev_op}, the linear solves have a relative tolerance of $10^{-12}$ in exact deflation while they stop at $10^{-3}$ when computing vectors for inexact deflation. The results are presented in \cref{tab:results_all}.

\begin{table}[h!]
\centering
\begin{tabular}{ |c|c|c|c|c|c| } 
\hline
Method & Deflation type & Nr. defl. vectors & eig.+direct & trace & total \\
\hline
MGMLMC & none & - & - & 276,826.66 & 276,826.66 \\
\hline
MGMLMC+skip & none & - & - & 130,208.2 & 130,208.2 \\
\hline
\multirow{4}{7em}{\begin{center}Hutchinson\end{center}} & \multirow{4}{5em}{\begin{center}exact\end{center}} & 1,024 & 595.51 & 102,838.06 & 103,433.57 \\ 
& & 2,048 & 3,337.336 & 102,060.58 & 105,397.916 \\ 
& & 4,096 & 11,597.91 & 78,727.95 & 90,325.86 \\ 
& & 8,192 & 77,765.96 & 74,980.76 & 152,746.72
 \\ 
\hline
\multirow{3}{7em}{MGMLMC+skip} & \multirow{3}{5em}{\begin{center}exact\end{center}} & 512 \& 510 & 3,633.41 & 9,903.29 & 13,536.70 \\ 
& & 1,024 \& 510 & 5,878.38 & 7,619.27 & 13,497.65 \\ 
& & 2,048 \& 510 & 13,731.51 & 7,359.11 & 21,090.62 \\ 
\hline
\multirow{3}{7em}{MGMLMC+skip} & \multirow{3}{5em}{\begin{center}inexact\end{center}} & 512 \& 510 & 2,274.88 & 10,073.90 & 12,348.78 \\ 
& & 1,024 \& 510 & 4,378.46 & 7,588.69 & 11,967.15 \\ 
& & 2,048 \& 510 & 9,959.23 & 7,266.92 & 17,226.15 \\ 
\hline
\end{tabular}
\caption{Execution times (in seconds) when computing $\mathrm{tr}(D^{-1})$ via various methods described in earlier sections.  In the third column, which displays the number of deflation vectors used, MGMLMC+skip has only two difference levels due to skipping the second one. The fourth column presents the times to obtain the deflation vectors and in the case of inexact deflation it includes the extra time due to inversions (see the paragraph right before \cref{eq:projector_inex_defl} on the need for these inversions), while the fifth column corresponds to the times for computing $\mathrm{tr}(D^{-1})$. The last column shows the total execution time i.e.\ for eigensolving plus computation of $\mathrm{tr}(D^{-1})$. }
\label{tab:results_all}
\end{table}

Instead of a cost model based on FLOPS, as in \cite{frommer2022multilevel}, we have opted for time measurements here. To reduce overheads due to the interpreter in Python, we measure execution times of very specific operations: $D_{\ell}$, deflations, $P_{\ell}$, $P_{\ell}^{H}$ and axpy operations.

As has already been reported in \cite{frommer2022multilevel} and displayed here in \cref{fig:results_mlmlc_paper}, exactly deflated Hutchinson clearly outperforms MGMLMC when $m_{0} = -0.132$. We revert this behaviour here via deflation on MGMLMC, as illustrated in \cref{tab:results_all}. In that table, MGMLMC+skip represents skipping the second difference level, which is of great computational benefit because in the multigrid hierarchy for the matrix used here, the number of nonzero elements in $D_{1}$ and $D_{2}$ are roughly the same. Skipping the second difference level impacts negatively the variance, and the sample size needed for convergence on the difference level going from $\ell=1$ to $\ell=3$ increases with respect to the original two levels when no skipping is done, but this increase in the sample size is small enough that we still see a large gain when skipping a level, as can be seen from the first two rows in \cref{tab:results_all}.

Skipping a level is not enough for MGMLMC to outperform exactly deflated Hutchinson, hence we resort to deflation on MGMLMC. As stated before, when eigensolving for exact deflation in the MGMLMC case, i.e.\ when computing eigenpairs of the operator $J_{\ell}$ in \cref{eq:hermitian_diff_lev_op}, we use a tolerance of $10^{-9}$ for the eigensolver and of $10^{-12}$ for the solves appearing in $M_{\ell}$ in each call of $J_{\ell}$ which happens at each iteration of the eigensolver. We can see in \cref{tab:results_all} that, for exactly deflated MGMLMC, the growth in execution time stops being linear at some point, and this is due to asking for a tolerance $10^{-9}$ in the eigensolver, which leads to the orthogonalizations within the eigensolver starting to dominate. This is less pronounced for inexactly deflated MGMLMC, where we ask for a tolerance of $10^{-1}$ from the eigensolver.

This dominance of the orthogonalizations in the eigensolver is expected when a very large number of deflation vectors is set, and we can see it already in exactly deflated Hutchinson in \cref{tab:results_all}. The size of the L3 cache in the machine where we performed our numerical experiments is 55 MB. A matrix of size $32,768 \times 128$ in double precision, which is the case of the matrix containing the deflation vectors when we deflate 128 of them, has a size of 64 MB, and this is too large to sustain coherence with the largest cache level. When using 128 deflation vectors, though, the multigrid solves are still considerably more expensive than the projections associated to deflation, especially because the Python installation that we use has been compiled with BLAS enabled, and we get BLAS3 performance for these projections.

These projections affect not only the eigensolver, but also the computation of the trace. In \cref{tab:results_all}, in the case of exactly deflated Hutchinson, we see practically no gain when going from 1,024 to 2,048 deflation vectors, stemming from the deflations going from 11.5\% to 18.7\% of the overall execution time of the trace, respectively. The same happens in deflated MGMLMC when going from 1,024 \& 510 to 2,048 \& 510, regardless of the type of deflation.

The best result is, as expected, MGMLMC with the combined use of skipping a level difference plus inexact deflation. With 1,024 \& 510 deflation vectors, this method outperforms the best case of exactly deflated Hutchinson by a factor of 7.5 in the overal execution time. This gain comes from the need of less deflation vectors for reducing the variance of each level difference, this coming in turn from the inexact deflation already performed by those difference levels. Furthermore, having less deflation vectors benefits the performance of the method computationally as well, as we have previuosly described.




\section{Outlook}

A first possible improvement for deflated multigrid multilevel Monte Carlo is to switch to an eigensolver more in accordance with the eigenproblem at hand. As described in \cref{sect:exact_defl_diff_levels}, the computation of the singular vectors is done via eigensolving with the operator $J_{\ell} = Q^{-1}_{\ell} - P_{\ell} Q^{-1}_{\ell+1} P^{H}_{\ell}$. This could be converted into a generalized eigenvalue problem with a non-Hermitian operator, which can then be treated via e.g.\ Jacobi-Davidson in a perhaps more efficient manner. We are currently working on implementing and testing inexactly deflated MGMLMC in the context of lattice QCD.

\bibliographystyle{unsrtnat}
\bibliography{references}

\end{document}